\documentclass{proc-a4}

\usepackage{lineno}
\usepackage{comment}
\usepackage{color}
\usepackage{doi}
\usepackage{soul}
\usepackage{graphicx}
\usepackage{todonotes}
\usepackage{subcaption}
\usepackage{amsmath}
\usepackage{amsfonts}

\begin{document}

\author{A. Settimi $\dag$}
\aff{Laboratory for Timber Construction (IBOIS, EPFL)}

\author{D. Gilliard $\dag$}
\aff{Laboratory for Timber Construction (IBOIS, EPFL)}

\author{E. Skevaki $\dag$}
\aff{Laboratory for Creative Computation (CRCL, EPFL)}

\author{M. Kladeftira}
\aff{Laboratory for Creative Computation (CRCL, EPFL), Structural Xploration Lab (SXL, EPFL)}

\author{J. Gamerro}
\aff{Independent Researcher}

\author{S. Parascho}
\aff{Laboratory for Creative Computation (CRCL, EPFL)}

\author{Y. Weinand}
\aff{Laboratory for Timber Construction (IBOIS, EPFL)}

\author{}
\aff{$\dag$ these authors contributed equally to this work}

\abstract{
In digital timber construction, scanning technologies and point cloud data are widely used due to the accessibility of affordable 3D sensors, photogrammetry, and user-friendly CAD tools. While typically not employed for accuracy checks in timber fabrication due to the precision of standard machinery, experimental research and prototyping with joinery and assembly can benefit from precision and accuracy evaluation tools.

We introduce diffCheck, a C++/Python software integrated into Grasshopper to address this need. It uses advanced point cloud analysis to compare scans of fabricated timber structures with their respective CAD models, helping to identify discrepancies. Tested on various timber elements and digital fabrication methods like robotic assembly, AR-assisted woodworking, and CNC machining, diffCheck aims to establish a user-friendly benchmark framework for digital fabrication systems using timber components, with the potential to find applications in other materials. Its source code and the analyzed data are openly shared with the digital fabrication community under a permissive license.
}


\chapter{DiffCheck: a Scan-CAD Evaluation Tool for Digital Manufacturing and Assembly Processes in Timber Construction}

%%%%%%%%%%%%%%%%%%%%%%%%%%%%%%%%%%%%%%%%%%%%%%%%%%%%%%%%%%%%%%%%%%%%%%%%%%%%%%%%%%%%%%
\section{Introduction}
%%%%%%%%%%%%%%%%%%%%%%%%%%%%%%%%%%%%%%%%%%%%%%%%%%%%%%%%%%%%%%%%%%%%%%%%%%%%%%%%%%%%%%

%% current context of proliferation of scans and data but processing and analysis tools are still needed for timber construction
In recent years, the availability of 3D scanning technologies has multiplied, with costs decreasing and accessibility improving. This proliferation of scanning tools has led to a significant increase in the amount of data that can be captured from physical objects. While the acquisition of scans has become easier, the processing of such raw scan data into useful insights for quality assessment remains complex.
%% fragmented pipelines to evaluate artifacts
Existing workflows often require users to navigate through multiple software platforms: one to generate CAD models, another to clean and process scans, and yet another to perform the comparison itself. This fragmented process can be cumbersome and time consuming for everyday digital makers and advanced users. Hence, we identify a lack of a unified platform that can streamline these tasks into a cohesive and user-friendly pipeline.
%% there are specifics to the evaluation of timber fabrication
In addition, despite the value of a generic scan-to-model comparison, timber fabrication has unique requirements for scan-to-CAD evaluations. Unlike other materials, timber construction often involves irregular shapes, organic growth patterns, and complex joinery, all of which demand specialized metrics to compare scans with digital models. A one-size-fits-all approach might not be sufficient for assessing timber fabrication, requiring tailored benchmark systems.
%% what we propose to fill these research gaps: diffcheck
A unified platform designed for digital makers in timber construction could significantly enhance the accessibility of these evaluations among digital fabrication developers. Hence, we propose diffCheck (DF), a C++/Python powered plug-in for Grasshopper that allows the user to identify discrepancies across point clouds and 3D models of individually machined timber pieces featuring various joints and fully assembled timber structures. The proposed software aims to make easy-to-use benchmarking software accessible for fabrication processes involving timber structures. DF provides online documentation [\cite{diffCheckDocu2024}], and its code is open source and shared with the community [\cite{diffCheckDataset2024}].

%% brief intro of the paper's structure
In the following, we present a comprehensive review of current scan-to-CAD evaluation methodologies for timber constructions, along with an overview of existing software tools for scan post-processing and manipulation. We then introduce the architecture and key functionalities integrated into DF. Next, two experimental DF's applications involving real-world timber digital construction scenarios are introduced, focusing on both additive (i.e. robotic or manual assembly) and subtractive processes (i.e., cutting and milling). Finally, we outline potential future developments for the tool, including expanding its functionality to support more complex fabrication workflows.

%%%%%%%%%%%%%%%%%%%%%%%%%%%%%%%%%%%%%%%%%%%%%%%%%%%%%%%%%%%%%%%%%%%%%%%%%%%%%%%%%%%%%%
\section{Related works}
%%%%%%%%%%%%%%%%%%%%%%%%%%%%%%%%%%%%%%%%%%%%%%%%%%%%%%%%%%%%%%%%%%%%%%%%%%%%%%%%%%%%%%
%% general intro
The state-of-the-art review is divided into two paragraphs, each addressing key research areas to which the DF aims to contribute. The first focuses on works related to fabrication evaluation in timber construction via scanning. In contrast, the second examines the open-source CAD-integrated point cloud processing plug-ins and free software available to this date. 

\subsection{Timber fabrication evaluation}
\label{sec:rel:timber_eval}
%% scanning in quality control for timber fabrication
Timber fabrication is not new to the use of scanning and sensing systems to ensure quality control.
%% glulam (since there was the most literature)
Free-form glulam production, similar to round wood manufacturing, is hindered by the lack of clear reference points. This called for early adoption of scanning techniques and point cloud, often live, processing to better align virtual models with the physical production of such elements. Consequently, this has led to the development of several point cloud-based sensing workflows [\cite{Svilans2019, Svilans2021, Vestartas2020, Larsen2020}].
%% subtractive
Digital assessment through scan-CAD can gauge differences between robotic milling of timber joinery against standard CNC tools [\cite{Pantscharowitsch}], or whenever a 3D model exists for any manually worked element.
%% robotic assembly
Concerning robotic manufacturing and assembly, scan-to-CAD comparison methodologies can be employed to evaluate the accuracy of the process, as errors can occur due to specific material behavior or operations [\cite{Ruan2023}], human agents [\cite{Skevaki2024}], or large scale prototyping [\cite{Mesnil2023}].

\subsection{Existing scan processing tools}
\label{sec:rel:timber_eval}
%% tools in Rh/Gh
Rhino's Grasshopper (GH) environment has recently been populated with multiple scan processing plug-ins. Point Cloud Components [\cite{Lin2014}] is one of the very first to propose point cloud processing and geometry-to-scan preliminary tools originally for landscape applications. Volvox [\cite{Zwierzycki2016}] proposes more advanced methods such as cropping, merging, and sub-sampling, and it tackles generic scan-CAD comparisons for some study cases at the building scale. It is also worth mentioning Tarsier [\cite{Tarsier2016}], a small-scale open-source project primarily utilized for point cloud visualization from sensing devices. Cockroach [\cite{SettimiCkrch2022}] is the first umbrella plug-in regrouping multiple external point cloud processing libraries, presenting an extensive, yet, as the rest of the mentioned plug-ins, very generic post-processing tool-set.
% outside CAD
Beyond the limitations of CAD, two of the most widely recognized free solutions for scan processing and comparison are CloudCompare [\cite{CloudCompare2016}] and MeshLab [\cite{MeshLab2008}]. Both provide an extensive range of functionalities, with CloudCompare being particularly adept at detecting discrepancies between point clouds and other data types, such as meshes or between point clouds. While these tools offer flexibility through plug-in systems and Python wrappers, they lack seamless integration within CAD environments. When switching between multiple software, we lose crucial semantic information necessary to track specific features of our timber components, such as joint faces or individual assembly elements, which are important in the fabrication process.

%%%%%%%%%%%%%%%%%%%%%%%%%%%%%%%%%%%%%%%%%%%%%%%%%%%%%%%%%%%%%%%%%%%%%%%%%%%%%%%%%%%%%%
\section{Methodology}
%%%%%%%%%%%%%%%%%%%%%%%%%%%%%%%%%%%%%%%%%%%%%%%%%%%%%%%%%%%%%%%%%%%%%%%%%%%%%%%%%%%%%%
%% quick intro
DiffCheck (DF) is a Grasshopper (GH) plug-in, built on a C++ core library, designed to detect discrepancies between point clouds and 3D models of both individually machined timber components with joints and fully assembled timber structures. The source code of the presented software is open, and accessible under a GPL-3.0 license in its public repository [\cite{diffCheckV12024}]. This chapter overviews DF's software architecture, underlying philosophy, and key functionalities.

\subsection{Software architecture}

%% The intro of the general architecture: 3 sections
The software architecture of DF is organized into three main sections. The first foundational portion of the source code is represented by a C++ umbrella library regrouping low-level dependencies such as Open3d [\cite{Zhou2018}], CGAL [\cite{cgal2024}] and Cilantro [\cite{Zampogiannis2018}] which power the back-end of the most complex and demanding computational functionalities DF can offer. This portion is wrapped into the second DF's component: an API written in Python 3.9.1 and distributed via PyPI (Python Package Index). Finally, the GH Python-based plug-in represents only the top-level visual scripting DF interface.

\begin{figure}
    \centering
    \includegraphics[width=120mm]{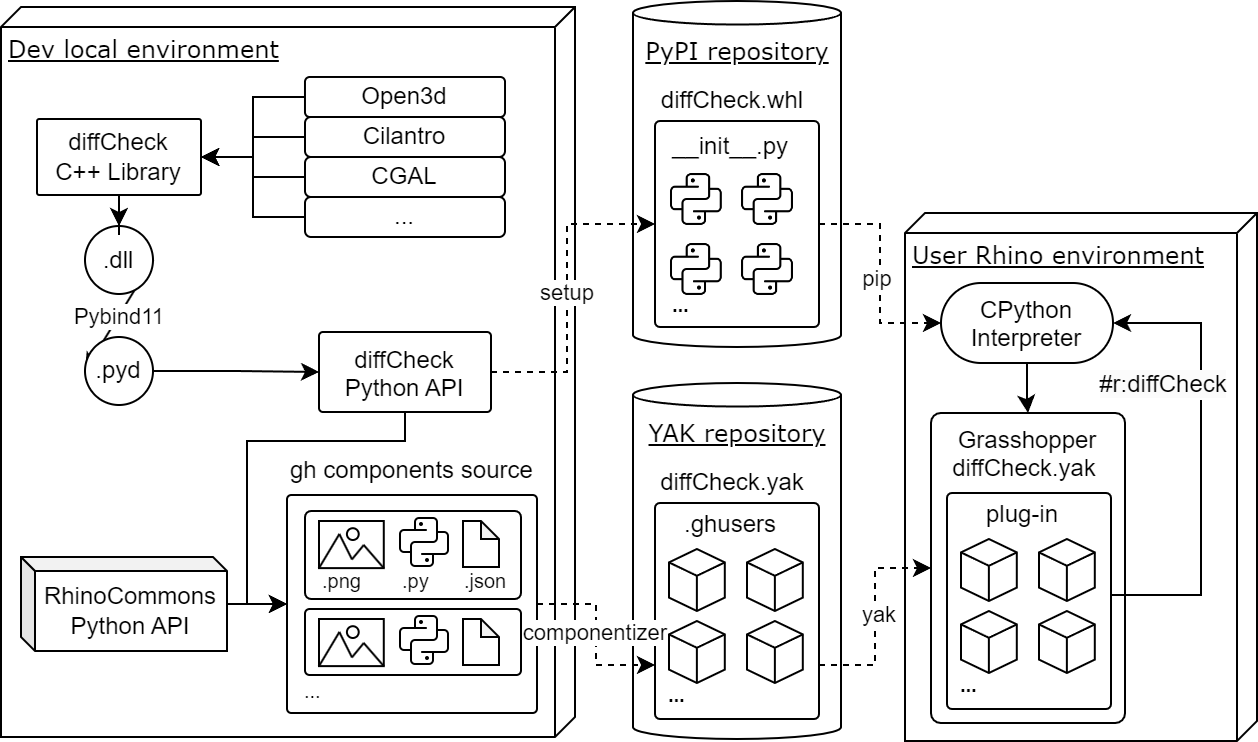} 
    \caption{A data flow diagram illustrating the composition and distribution of DiffCheck. Leveraging Rhino’s CPython interpreter, most of the source code is modularly developed and distributed, enabling a streamlined, user-friendly installation process within Grasshopper.}
    \label{fig:metho:softarch}
\end{figure}

%% GH Python plug-in anatomy
With the release of Rhino v.8 [\cite{rhinoceros2023}], which integrates CPython into its .NET ecosystem, we developed a fully Python-based Grasshopper (GH) plug-in. Distributed via Rhino’s official Yak manager, the plug-in requires no additional installations from the user. Similar to the Compas framework [\cite{compas2019}], the plug-in's components are .ghuser objects. This design offers significant advantages for continuous integration (CI) practices, enabling automatic documentation for each component. Moreover, we believe that a full-fledged Python Grasshopper plug-in increases the likelihood of broader contributions. Python's simplicity and widespread use make it more accessible to a larger public. 
The generated and distributed components are flagged for Rhino’s CPython interpreter to automatically download and use a specific version of the DF PyPI package's relevant version. This approach allows most of the plug-in's source code to be shipped as a standalone Python package, resulting in a more lightweight, modular, and user-friendly installation while decoupling the development and distribution environment as shown in the scheme in Fig.~\ref{fig:metho:softarch}.

\subsection{diffCheck's functionalities}

%% the base of DF is a general 
We have designed the plugin with a modular structure to make DF broadly applicable and allow users to customize their evaluation pipelines. Users can tailor their workflows by assembling and combining different components to fit specific project needs. For the most part, DF operates as a general-purpose point cloud processing tool, offering a suite of functions that can be applied to a wider range of applications outside the currently proposed scope. Some key functions available in DF include removal of point cloud outliers, down-sampling, registrations and clustering, brep- or mesh-to-cloud sub-sampling, and point cloud I/O utilities. 
%% DF data structures to represent timber structures
This very first flexible foundation of DF is complemented by specific data substructures designed to represent the intricacies of timber structures at different scales. At the beginning of every process, the entire CAD model is parsed into a \textit{DFAssembly} object, which registers all the distinct elements (\textit{DFBeam}), detects all their joints (\textit{DFJoint}), and terminates by refining the joint's structure into its composing faces (\textit{DFFace}).
%% computational component
In addition to the modular components for analyzing assemblies and joints, DiffCheck includes a dedicated error computation tool. This specific component is designed to compute and quantify discrepancies between the point cloud and DF's model objects, providing detailed error metrics. Key functionalities of this component include overall error calculations, threshold settings, error categorization, and heatmap visualization. This error computation tool ensures that users can thoroughly evaluate the precision of their projects, offering clear, actionable insights within the CAD environment.

%% intro into the study-cases section
By leveraging modular point cloud processing components, internal data structures, and precise error analysis, DF provides tailored workflows that assess the accuracy and quality of timber fabrication at various stages. The following section presents a selection of study cases where DF is used to create specific evaluation pipelines for assembly and subtractive processes (e.g. cutting or milling) in timber construction.

%%%%%%%%%%%%%%%%%%%%%%%%%%%%%%%%%%%%%%%%%%%%%%%%%%%%%%%%%%%%%%%%%%%%%%%%%%%%%%%%%%%%%% 
\section{Evaluation}
%%%%%%%%%%%%%%%%%%%%%%%%%%%%%%%%%%%%%%%%%%%%%%%%%%%%%%%%%%%%%%%%%%%%%%%%%%%%%%%%%%%%%%
In this chapter, we provide two relevant study cases of digital manufacturing and assembly processes in timber construction that demonstrate how DF can provide insights to evaluate the resulting fabricated elements.
 The employed point clouds are obtained using a hand-held solid-state infra-red LiDAR scanner (FARO\textregistered Freestyle2) with resolutions close to ~0.5 mm. All the evaluation material is made accessible via a public Zenodo repository [\cite{diffCheckDataset2024}]. We also provide hands-on tutorials for each presented evaluation pipeline in our online documentation [\cite{diffCheckDocu2024}].

\subsection{Assembly studycase}

In this section, we present three case studies of non-standard timber assemblies and their accuracy evaluation with quantitative metrics provided by DF. To demonstrate the versatility of DF we will present: (a) a roof structure detail connected with half-lap scarf joints and half-lap cross joints that were manually fabricated with Augmented Reality (AR) assistance and assembled with the aid of standard construction tools and equipment, (b) a frame structure with four wood logs connected with half-lap cross joints, as well as (c) a 2-robot cooperatively assembled spatial structure with a total of thirteen elements of square section connected through bolted face lap joints. Two ABB GoFa CRB 15000-5 with a 0.95-meter reach were used for the robotic assembly. 
%% pre-processing (i.e. down-sampling)
Once the structure is assembled, a point cloud is acquired using the scanning device described above. Depending on the resolution of the scanning device of choice and the processing power of the computer used, we propose down-sampling the acquired point cloud with DF using a user-variable factor.

\begin{figure}
    \centering
    \includegraphics[width=150mm]{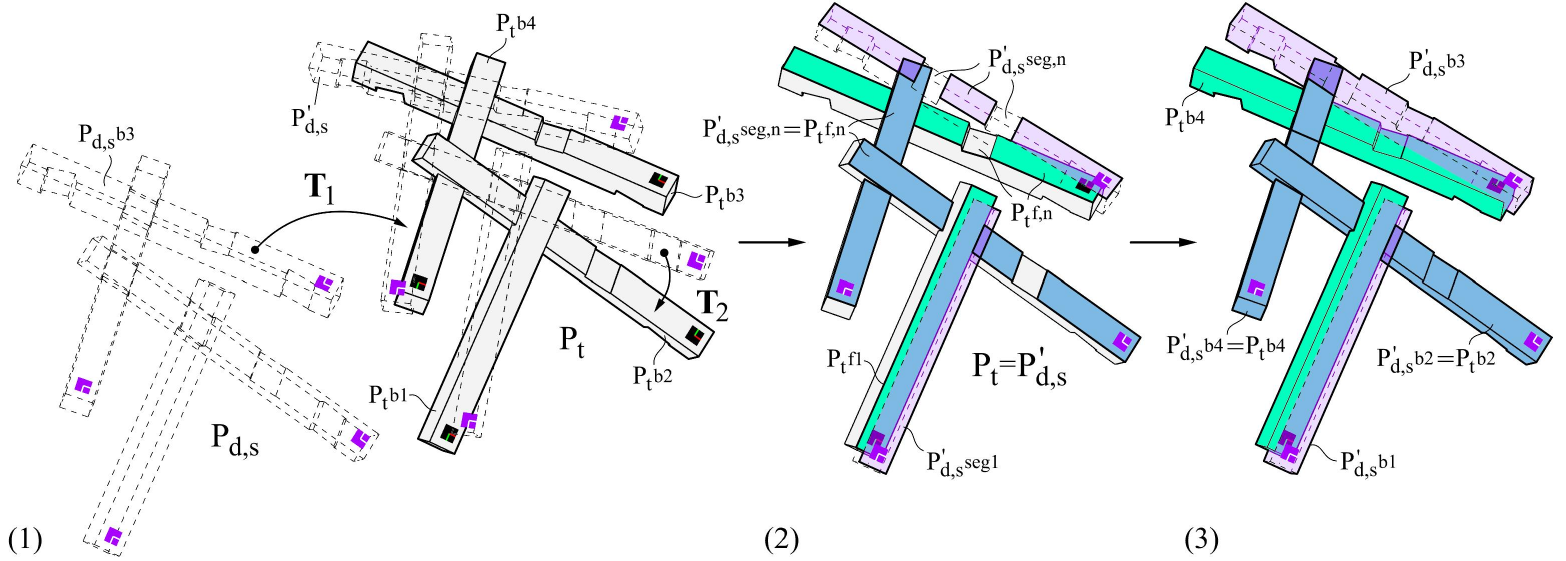} 
    \caption{Illustration of the assembly evaluation: (1) the ground truth $P_t$ and scanned $P_s$ point clouds are registered to the same reference system, (2) segmentation by normal and face association, (3) each beam is finally detected within the scan.}
    \label{fig:eval:addscheme}
\end{figure}

%% scan-CAD processing and error computation
Let $P_{d,s}$ represent the down-sampled, scanned source cloud, and $P_t$ represent the target point cloud sub-sampled from the CAD model. The first step of the evaluation process with diffCheck is to register $P_{d,s}$ to $P_t$. This initial rigid transformation $T_1: \mathbb{R}^3 \xrightarrow{} \mathbb{R}^3$ can be obtained by either a set of fiducial markers or a DF's RANSAC-based registration component. The transformation obtained can be further improved using an Iterative Closest Point (ICP) refinement (Fig.~\ref{fig:eval:addscheme}.1). With the two point clouds registered, $P'_{d,s}$ must now be segmented into clusters representing the individual beams in the structure. This segmentation enables a one-to-one comparison for each assembly element, ensuring that the displacement of each beam can be determined independently. The transformed source point cloud undergoes first a normal-based segmentation producing a given number of $P'_{d,s}{seg,n}$ representing the beams' faces. Subsequently, for each beam face $P_t{f}$ of $P_t$, the closest and most similarly oriented segment $P'{d,s}{seg}$ is identified (Fig.~\ref{fig:eval:addscheme}.2). Leveraging the geometric semantics of the CAD model, the detected faces $P'{d,s}{f,n}$ can then be mapped back to the corresponding beam $P_t{b}$. This process produces a set of point clouds, each representing an individual assembly element $P'{d,s}{b}$ (Fig.~\ref{fig:eval:addscheme}.3).
Once the raw scan and the CAD are pre-processed, we can evaluate the assembly error by computing a set of metrics based on distance values between each $P'{d,s}{b}$ and $P_t{b}$: the mean of point-to-point distances, the mean squared error, the standard deviation, as well as min/max deviation. These errors are visualized with an adaptive color gradient in DF's custom visualizer, either on $P_s$ or on the target CAD model.

\begin{figure}
    \centering
    \includegraphics[width=150mm]{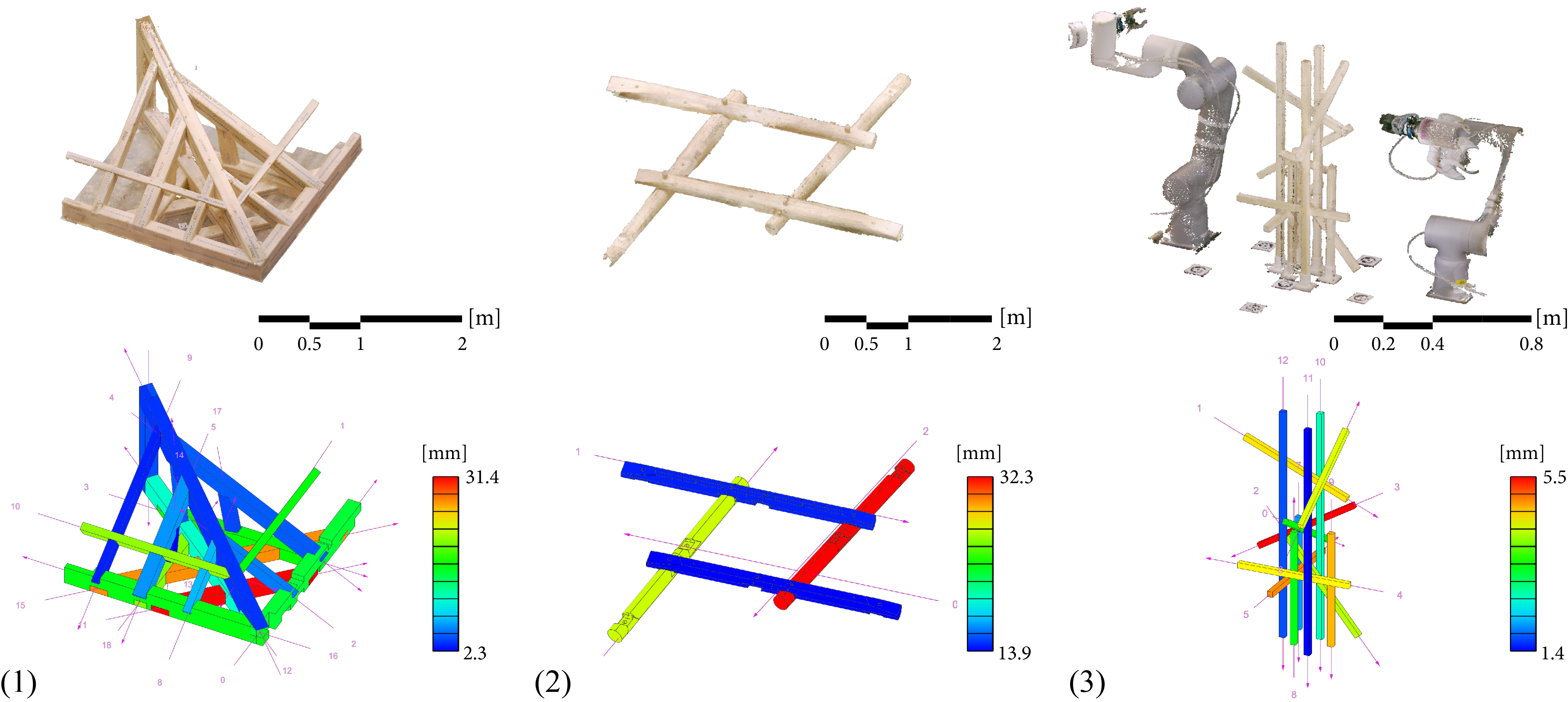} 
    \caption{Illustration of the mean error results per element on the CAD model for three assembly study cases: (1) roof structure segment crafted by augmented workers and assembled manually, (2) frame structure with four wood logs connected by CNC-fabricated half-lap cross joints, (3) spatial structure assembled by two robots and a human.}
    \label{fig:additiveall}
\end{figure}

Implementing this evaluation pipeline for the three case studies (Fig.~\ref{fig:additiveall}) we observe a mean member deviation between 3.7±1.3 mm for the robotic assembly, 8.9±7.9 mm for the roof segment and 21.8±7.9 mm for the log frame. These measured errors should be understood as assembly-specific metrics that can give the user further insights into the fabrication process and not as a 1:1 comparison between the different case studies as they vary substantially in scale, section, and connecting logic. For instance, we can observe that the robotic assembly (Fig.~\ref{fig:additiveall}.3) has an overall lower mean error. However, the results of DF's evaluation can also be used to identify discrepancies in the placement between the two cooperating robots, small shifts that occurred while fastening performed by humans in difficult-to-reach positions, and other errors resulting from uncertain factors. Similarly, the roof assembly (Fig.~\ref{fig:additiveall}.1) shows an overall low mean error with two outlier elements with a mean error approx. 10x higher than the median. This can be explained by a sizing discrepancy as these two beams have a different section profile than their respective CAD model. The error measured by DF in this case is a combined placement and sizing error. Nevertheless, the scan was correctly registered to the CAD model, and DF successfully captured this incoherence.

The evaluation pipeline remains largely similar in all three cases of assembly, with only a few differences. Specifically, in the robotic assembly the global registration can be omitted in place of the use of fiduciary markers because of the known transformation between the virtual and real world. At the same time, in the other scenarios, this is not the case. Therefore, DF's \textit{RANSAC Global registration} should first be applied to achieve a coarse alignment with $P_t$ and use that transformation to obtain $P'{d,s}$. Furthermore, in the case of round timber logs (Fig.~\ref{fig:additiveall}.2), such irregular geometries are more challenging to approximate in digital space, hence we expect higher errors both in the registration and error calculation of the assembly. It is worthwhile to note that the precision in assembly in such cases, as in all structures that feature wood-wood connections, relies heavily on the manufacturing precision of the joints, therefore an evaluation of the individual joint precision is of relevance as shown in the following section.

%% ------------------------------------------------------------------
%% ------------------------------------------------------------------
\subsection{Subtractive study case}
\begin{figure}
    \centering
    \includegraphics[width=140mm]{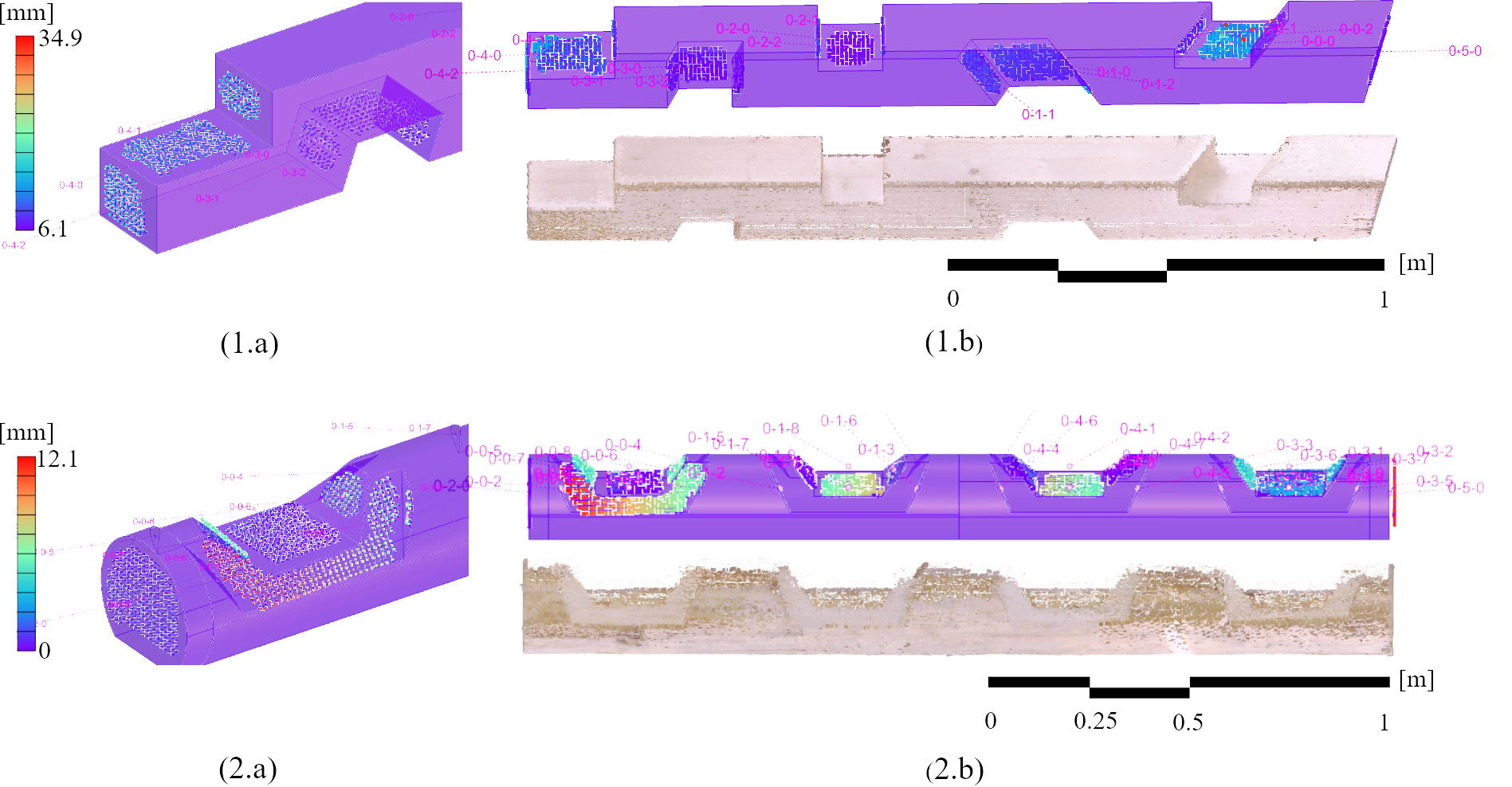} 
    \caption{Screen capture of the Rhino workspace. (1.a) and (2.a): close-ups on single joints, (1.b) and (2.b) bottom: scan of the evaluated piece, above: 3d model and highlighted scan evaluation.}
    \label{fig:eval:scancadsub}
\end{figure}

%% brief intro of the section
This section presents two subtractive study case sets where DF can inform and provide quantitative insights into joints rather than the assembly. The first case study is a comparison between 3 squared section beams of ~2 m length (Fig.~\ref{fig:eval:scancadsub}.1) containing each 1 half-lap, 4 half-lap cross, and 1 butt joints machined with three different fabrication techniques: AR-guided manual circular saw and chainsaw cutting, and a 5-axis CNC router (MAKA System BC 170). The second experiment is meant to broaden the range of timber shapes that DF can accurately assess. Hence, we include the analysis of 1 round wood beam (Fig.~\ref{fig:eval:scancadsub}.2) fabricated by the same CNC milling machine. This highlights a challenging scenario where the 3D model cannot be perfectly replicated in reality due to the irregularity of the tree trunk.

%% analysis methodology for joints
DF enables analysis of joints at two distinct levels: i) \textit{per-joint}, where it is the joint's location to be evaluated within the timber element, and ii) \textit{per-joint-face}, in which the joint's faces are evaluated. These two levels correspond to two different types of errors. The first helps assess the overall positioning of the joint, while the second informs the accuracy of execution of each boundary cut of the joint.

\begin{figure}
    \centering
    \includegraphics[width=150mm]{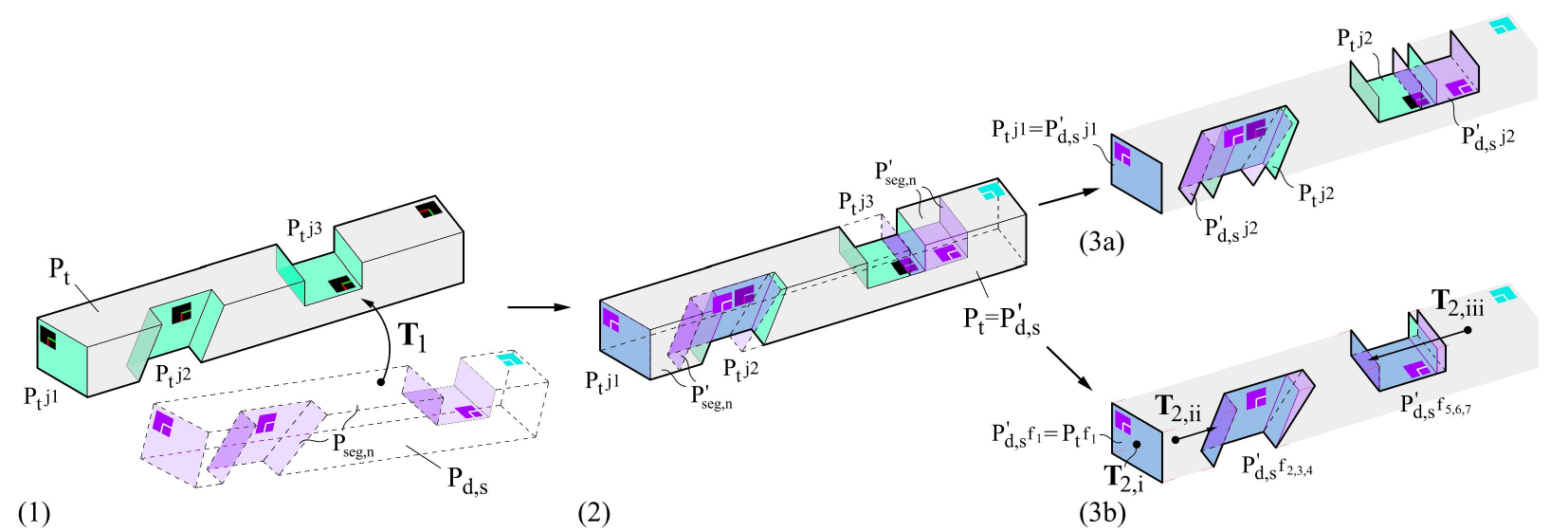} 
    \caption{Representation of the \textit{per-joint} and \textit{per-joint-face} evaluations: (1) the ground truth $P_t$ and scanned $P_s$ point clouds are registered to the same reference system, (2) segmentation phase, (3a) per-joint, and (3b) per-joint-face processing.}
    \label{fig:eval:subscheme}
\end{figure}

%% subtractive preparation: registration and segmentation
As in the additive study cases, the evaluation pipeline commences by down-sampling ($P_s \xrightarrow{down} P_{d, s} $) and aligning it to the corresponding sub-sampled point cloud of the 3D model ($P_t$):$P'_{d, s} = T_1(P_{d, s})$ as illustrated in Fig.~\ref{fig:eval:subscheme}.1. DF provides RANSAC-based and ICP registrations [\cite{Zhou2018}] that calculate this first transformation $T_1: \mathbb{R}^3 \xrightarrow{} \mathbb{R}^3 $. Next, $P'_{d, s}$ is segmented based on the normals resulting in coherent point cloud segments presenting similar normal orientation $\mathbf{P'}_{seg} = P'_{seg, 1}, P'_{seg, 2}, \ldots P'_{seg, n}$ (Fig.~\ref{fig:eval:subscheme}.2).
%% per-joint evaluation pipeline
For the \textit{per-joint} analysis (Fig.~\ref{fig:eval:subscheme}.3a), the closest and most similarly oriented $P'_{seg}$ is found for each $P_t$'s joint face $P_t{f}$. The points of $P'_{seg}$ that are perpendicularly projectable onto $P_t{f}$ within a user-defined tolerance, are then grouped into one point cloud $P'_{d, s}{f}$. All the $P'_{d, s}{f, n}$ belonging to the same joint are grouped into a single point cloud $P'_{d, s}{j}$. This produces an ensemble of $P'_{d, s}{j}$ that can be directly compared with the 3D model's target joint $P_{t}{j}$ to derive per-joint metrics.
%% per-joint-face evaluation pipeline
To obtain the \textit{per-joint-face} metrics (Fig.~\ref{fig:eval:subscheme}.3b), we employ $P'_{d, s}{j}$, obtained during the previous step, to calculate a single transformation $T_2: \mathbb{R}^3 \xrightarrow{} \mathbb{R}^3$ between $P'_{d, s}{j}$ and $P_t{j}$. This transformation is then applied to each $P'_{d, s}{f, i}$ associated with $P'_{d, s}{j}$. This step eliminates any overall displacement between the joint's ground truth and the corresponding scan segments. Eventually, the error distances between $P'_{d, s}{f}$'s points and the corresponding $P_t{f}$ can be computed to obtain per-joint-face values.

\begin{figure}
    \centering
    \includegraphics[width=130mm]{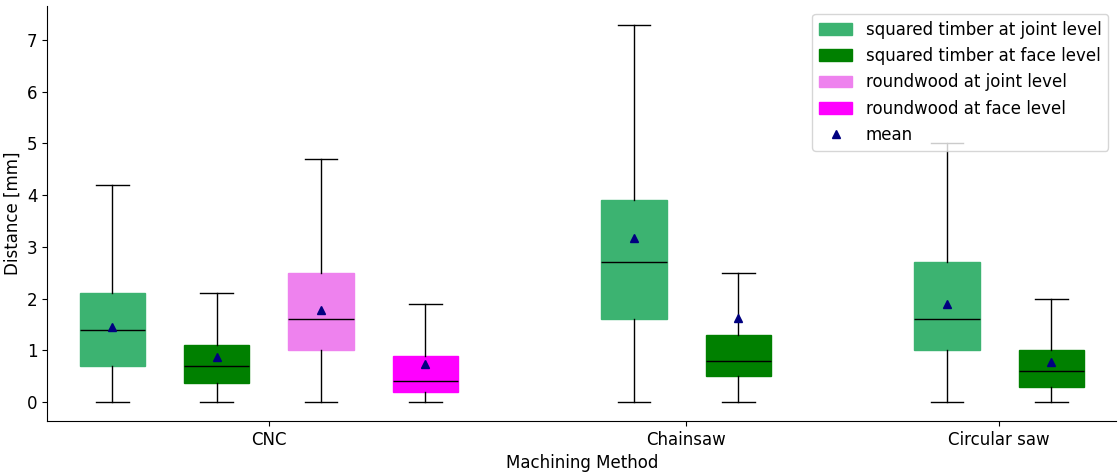} 
    \caption{DF's analysis at joint and joint's face level on a total of 24 joints composed by 70 individual faces. }
    \label{fig:eval:joint_plot}
\end{figure}

%% results and discussion at per-joint level
The results produced by the proposed pipelines are illustrated in Figure \ref{fig:eval:joint_plot}. As it could be expected from a fabrication technique standpoint, DF's error analysis reflected the accuracy of the tools by quantifying it: CNC-cut joints were the closest to the 3D model with an error value of 1.5±1.0 mm, directly followed by the AR-guided circular saw (1.9±0.8 mm), while chainsaw cuts had the largest deviation from the 3D model (3.2±3.1 mm).
Additionally, DF highlighted the best-executed joint among the four beams, which was the CNC-cut squared element with a 1.1±0.8 mm, while the less accurate joint belongs to the chainsaw-cut element (5.1±5.2 mm).
%% results and discussion at per-joint-face level
Considering the \textit{per-joint-face evaluation} (Fig. \ref{fig:eval:joint_plot}), the mean error for the face cut with the CNC, circular saw, and chainsaw were 0.9±0.8 mm, 0.8±0.8 mm, and 1.6±3.2 mm respectively. Overall, among the faces evaluated by DF on this dataset, the best ones belong to the CNC-cut squared and round wood elements (0.3±0.7 mm and 0.3±0.3 mm respectively). The most defective cut belonged to chainsaw cutting with an error close to 12.7±6.1 mm. Moreover, DF’s benchmark highlighted a shared trend among all the examined fabrications: the positioning of the joints exhibits a greater error compared to the execution of the joints themselves.
%% results for round wood
Finally, in the case of the CNC-cut round wood, DF was able to analyze the joints and produced results comparable to the squared beam cut with the CNC (1.8±1.1 mm), demonstrating a capacity to detect and analyze joints in a more organic shape, such as a round wood beam, by filtering the noise produced by e.g., the log's irregular portions of the bark.

%%%%%%%%%%%%%%%%%%%%%%%%%%%%%%%%%%%%%%%%%%%%%%%%%%%%%%%%%%%%%%%%%%%%%%%%%%%%%%%%%%%%%% 
\section{Conclusions}
%%%%%%%%%%%%%%%%%%%%%%%%%%%%%%%%%%%%%%%%%%%%%%%%%%%%%%%%%%%%%%%%%%%%%%%%%%%%%%%%%%%%%%

%% quick recap and highlight of the achieved goal
We have introduced an open-source and user-friendly software solution that simplifies and democratizes the processing of scan data to evaluate timber structures, thereby assessing their fabrication processes. By providing a set of documented and accessible pipelines in visual scripting, based on real-scale timber structures and mock-ups, we guide entry-level users through evaluating digital processes in timber construction both in additive and subtractive domains. The modular and flexible design of DF, paired with its specialized tools for assembly and joint analysis, can promote reproducibility and accessibility in open science practices for digital fabrication. Given the interdisciplinary nature of digital construction, which includes computer science, human sciences, and construction heuristics, it's often difficult to allocate adequate resources for thorough scientific evaluation. This tool seeks to address those challenges and streamline the evaluation process.

%% future avanues
Looking ahead, we aim to enhance the software’s capabilities and extend its use to a wider array of timber structures. For example, the integration of finer-grained features, such as precise handling of holes and curved shapes, is not yet fully realized in the current version. Additionally, future research could explore adapting the DF’s tool-set for real-time feedback in robotic fabrication, paving the way for intelligent processes and seamless benchmark data integration. This could broaden its appeal to a larger community of robotic system developers and users, expanding its impact across the digital fabrication landscape.

%%%%%%%%%%%%%%%%%%%%%%%%%%%%%%%%%%%%%%%%%%%%%%%%%%%%%%%%%%%%%%%%%%%%%%%%%%%%%%%%%%%%%% 
%% References

%%%%%%%%%%%%%%%%%%%%%%%%%%%%%%%%%%%%%%%%%%%%%%%%%%%%%%%%%%%%%%%%%%%%%%%%%%%%%%%%%%%%%% 


\begin{thebibliography}{10}
\providecommand{\url}[1]{#1}
\csname url@samestyle\endcsname
\providecommand{\newblock}{\relax}
\providecommand{\bibinfo}[2]{#2}
\providecommand{\BIBentrySTDinterwordspacing}{\spaceskip=0pt\relax}
\providecommand{\BIBentryALTinterwordstretchfactor}{4}
\providecommand{\BIBentryALTinterwordspacing}{\spaceskip=\fontdimen2\font plus
\BIBentryALTinterwordstretchfactor\fontdimen3\font minus \fontdimen4\font\relax}
\providecommand{\BIBforeignlanguage}[2]{{%
\expandafter\ifx\csname l@#1\endcsname\relax
\typeout{** WARNING: IEEEtran.bst: No hyphenation pattern has been}%
\typeout{** loaded for the language `#1'. Using the pattern for}%
\typeout{** the default language instead.}%
\else
\language=\csname l@#1\endcsname
\fi
#2}}
\providecommand{\BIBdecl}{\relax}
\BIBdecl

\bibitem{diffCheckDocu2024}
A.~Settimi, D.~Gilliard, E.~M. Skevaki, and M.~Kladeftira, ``{D}iff{C}heck: {C}{A}{D}-{S}can comparison online documentation,'' \url{https://diffcheckorg.github.io/diffCheck/}, 2024, [Accessed 21-09-2024].

\bibitem{diffCheckDataset2024}
\BIBentryALTinterwordspacing
A.~Settimi, E.~Skevaki, M.~Kladeftira, D.~Gilliard, J.~Gamerro, S.~Parascho, and Y.~Weinand, ``\BIBforeignlanguage{en}{Support material for diffcheck publication for icsa 2025},'' 2024. [Online]. Available: \url{https://zenodo.org/doi/10.5281/zenodo.13939225}
\BIBentrySTDinterwordspacing

\bibitem{Svilans2019}
\BIBentryALTinterwordspacing
T.~Svilans, M.~Tamke, M.~R. Thomsen, J.~Runberger, K.~Strehlke, and M.~Antemann, \emph{New Workflows for Digital Timber}.\hskip 1em plus 0.5em minus 0.4em\relax Springer International Publishing, 2019, p. 93–134. [Online]. Available: \url{http://dx.doi.org/10.1007/978-3-030-03676-8_3}
\BIBentrySTDinterwordspacing

\bibitem{Svilans2021}
\BIBentryALTinterwordspacing
T.~Svilans, ``Glulamb: A toolkit for early-stage modelling of free-form glue-laminated timber structures,'' in \emph{Proceedings of the 2021 European Conference on Computing in Construction}, ser. EC3 2021, vol.~2.\hskip 1em plus 0.5em minus 0.4em\relax University College Dublin, Jul. 2021, p. 373–380. [Online]. Available: \url{http://dx.doi.org/10.35490/EC3.2021.194}
\BIBentrySTDinterwordspacing

\bibitem{Vestartas2020}
\BIBentryALTinterwordspacing
P.~Vestartas and Y.~Weinand, ``Laser scanning with industrial robot arm for raw-wood fabrication,'' in \emph{Proceedings of the 37th International Symposium on Automation and Robotics in Construction (ISARC)}, ser. ISARC2020.\hskip 1em plus 0.5em minus 0.4em\relax International Association for Automation and Robotics in Construction (IAARC), Oct. 2020. [Online]. Available: \url{http://dx.doi.org/10.22260/ISARC2020/0107}
\BIBentrySTDinterwordspacing

\bibitem{Larsen2020}
\BIBentryALTinterwordspacing
L.~N. Martin and A.~A. Kruse, ``Robotic processing of crooked sawlogs for use in architectural construction,'' \emph{Construction Robotics}, vol.~4, no. 1–2, p. 75–83, May 2020. [Online]. Available: \url{http://dx.doi.org/10.1007/s41693-020-00028-7}
\BIBentrySTDinterwordspacing

\bibitem{Pantscharowitsch}
\BIBentryALTinterwordspacing
M.~Pantscharowitsch and B.~Kromoser, ``Robotic timber milling: accuracy assessment for construction applications and comparison to a cnc-machine,'' \emph{Wood Material Science \& Engineering}, vol.~0, no.~0, pp. 1--17, 2024. [Online]. Available: \url{https://doi.org/10.1080/17480272.2024.2366487}
\BIBentrySTDinterwordspacing

\bibitem{Ruan2023}
\BIBentryALTinterwordspacing
D.~Ruan, W.~Mcgee, and A.~Adel, ``Reducing uncertainty in multi-robot construction through perception modelling and adaptive fabrication,'' in \emph{Proceedings of the 40th International Symposium on Automation and Robotics in Construction}, ser. ISARC2023.\hskip 1em plus 0.5em minus 0.4em\relax International Association for Automation and Robotics in Construction (IAARC), Jul. 2023. [Online]. Available: \url{http://dx.doi.org/10.22260/ISARC2023/0006}
\BIBentrySTDinterwordspacing

\bibitem{Skevaki2024}
\BIBentryALTinterwordspacing
E.~Skevaki, M.~Kladeftira, A.~N. Pittiglio, and S.~Parascho, ``Assembly of spaceframes in hybrid teams: Adaptive digital fabrication workflows for human-robot collaboration,'' \emph{ACCELERATED DESIGN - Proceedings of the 29th CAADRIA Conference, Singapore}, vol.~3, p. 291–300, 2024. [Online]. Available: \url{https://papers.cumincad.org/cgi-bin/works/Show?caadria2024_282}
\BIBentrySTDinterwordspacing

\bibitem{Mesnil2023}
\BIBentryALTinterwordspacing
R.~Mesnil, T.~Gobin, L.~Demont, P.~Margerit, N.~Ducoulombier, C.~Douthe, and J.-F. Caron, ``Flexible digital manufacturing of timber construction: the design and fabrication of a free-form nexorade,'' \emph{Construction Robotics}, vol.~7, no.~2, p. 193–212, Jun. 2023. [Online]. Available: \url{http://dx.doi.org/10.1007/s41693-023-00105-7}
\BIBentrySTDinterwordspacing

\bibitem{Lin2014}
\BIBentryALTinterwordspacing
E.~S. Lin and C.~Girot, ``Point cloud components tools for the representation of large scale landscape architectural projects,'' 2014. [Online]. Available: \url{https://api.semanticscholar.org/CorpusID:219618472}
\BIBentrySTDinterwordspacing

\bibitem{Zwierzycki2016}
\BIBentryALTinterwordspacing
M.~Zwierzycki, H.~L. Evers, and M.~Tamke, ``Parametric architectural design with point-clouds - volvox,'' in \emph{Proceedings of the 34th International Conference on Education and Research in Computer Aided Architectural Design in Europe (eCAADe) [Volume 2]}, ser. eCAADe 2016, vol.~2.\hskip 1em plus 0.5em minus 0.4em\relax eCAADe, 2016, p. 673–682. [Online]. Available: \url{http://dx.doi.org/10.52842/conf.ecaade.2016.2.673}
\BIBentrySTDinterwordspacing

\bibitem{Tarsier2016}
\BIBentryALTinterwordspacing
C.~Newnham and J.~Gwyllim, ``Tarsier,'' [Accessed 22-09-2024]. [Online]. Available: \url{https://bitbucket.org/camnewnham/tarsier/src/master/}
\BIBentrySTDinterwordspacing

\bibitem{SettimiCkrch2022}
\BIBentryALTinterwordspacing
A.~Settimi, P.~Vestartas, J.~Gamerro, and Y.~Weinand, ``Cockroach: an open-source tool for point cloud processing in {CAD},'' in \emph{{POST{\textendash}CARBON} {\textendash} Proceedings of the 27th {CAADRIA} Conference}, N.~G.~J. van Ameijde, K.~H. Hyun, D.~Luo, and U.~Sheth, Eds., Apr. 2022, pp. 325--334. [Online]. Available: \url{https://doi.org/10.52842/conf.caadria.2022.2.325}
\BIBentrySTDinterwordspacing

\bibitem{CloudCompare2016}
\BIBentryALTinterwordspacing
``Cloudcompare (version 2.10.3),'' [Accessed 22-09-2024]. [Online]. Available: \url{from http://www.cloudcompare.org}
\BIBentrySTDinterwordspacing

\bibitem{MeshLab2008}
P.~Cignoni, M.~Callieri, M.~Corsini, M.~Dellepiane, F.~Ganovelli, and G.~Ranzuglia, ``{MeshLab: an Open-Source Mesh Processing Tool},'' in \emph{Eurographics Italian Chapter Conference}, V.~Scarano, R.~D. Chiara, and U.~Erra, Eds.\hskip 1em plus 0.5em minus 0.4em\relax The Eurographics Association, 2008.

\bibitem{diffCheckV12024}
\BIBentryALTinterwordspacing
A.~Settimi, D.~Gilliard, E.~M. Skevaki, and M.~Kladeftira, ``diffcheckorg/diffcheck: v1.0.0: diffcheck v1,'' 2024. [Online]. Available: \url{https://zenodo.org/doi/10.5281/zenodo.13843959}
\BIBentrySTDinterwordspacing

\bibitem{Zhou2018}
Q.-Y. Zhou, J.~Park, and V.~Koltun, ``{Open3D}: {A} modern library for {3D} data processing,'' \emph{arXiv:1801.09847}, 2018.

\bibitem{cgal2024}
\BIBentryALTinterwordspacing
{The CGAL Project}, \emph{{CGAL} User and Reference Manual}, {5.6.1}~ed.\hskip 1em plus 0.5em minus 0.4em\relax {CGAL Editorial Board}, 2024. [Online]. Available: \url{https://doc.cgal.org/5.6.1/Manual/packages.html}
\BIBentrySTDinterwordspacing

\bibitem{Zampogiannis2018}
\BIBentryALTinterwordspacing
K.~Zampogiannis, C.~Fermuller, and Y.~Aloimonos, ``cilantro,'' in \emph{Proceedings of the 26th {ACM} international conference on Multimedia}.\hskip 1em plus 0.5em minus 0.4em\relax {ACM}, Oct. 2018. [Online]. Available: \url{https://doi.org/10.1145/3240508.3243655}
\BIBentrySTDinterwordspacing

\bibitem{rhinoceros2023}
R.~McNeel \emph{et~al.}, ``Rhinoceros 3d, version 8.0,'' \emph{Robert McNeel \& Associates, Seattle, WA}, 2023.

\bibitem{compas2019}
\BIBentryALTinterwordspacing
T.~V. Mele \emph{et~al.}, ``{COMPAS}: A framework for computational research in architecture and structures.'' 2017-2019, http://compas.dev. [Online]. Available: \url{https://doi.org/10.5281/zenodo.2594510}
\BIBentrySTDinterwordspacing

\end{thebibliography}
\end{document}